\documentclass[article]{JHEP3}

\usepackage{amsmath,amssymb}
\usepackage[dvips]{graphics}
\usepackage{epsfig}
\usepackage{psfrag}
\usepackage{bm}
\usepackage{amsmath,amssymb,epsfig,float}
\newcommand{\lb}{\left(}
\newcommand{\rb}{\right)}

\newcommand{\be}{\begin{equation}}
\newcommand{\ee}{\end{equation}}
\newcommand{\bea}{\begin{eqnarray}}
\newcommand{\eea}{\end{eqnarray}}
\newcommand{\bean}{\begin{eqnarray*}}
\newcommand{\eean}{\end{eqnarray*}}

\preprint{{\small \texttt{}}}

\title{Entropy function and higher derivative corrections to entropies in (anti-)de Sitter space}
\author{ Fu-Wen Shu and Xian-Hui Ge \footnote{Fu-Wen Shu and Xian-Hui Ge contribute to this paper equivalently.}
\\
\\
Asia Pacific Center for Theoretical Physics, Pohang 790-784, Korea\\
\\
\email{E-mail: fwshu@apctp.org}, \quad \email{gexh@apctp.org} }

\abstract{\\ ~~~We first briefly discuss the relation between black
hole thermodynamics and the entropy function formalism. We find that
an equation which governs the relationship between Sen's entropy
function and black hole entropy, can quickly give higher order
corrections to entropy of pure (anti-) de Sitter space without
knowing the corrected metric. We also show that near horizon
geometry and the entropy function extremization is no longer
required for pure (anti-)de Sitter space. The entropy of (anti-)de
Sitter space and Schwarzschild-(anti-) de Sitter black holes
together with Gauss-Bonnet terms,  $R^2$ terms and $R^4$ terms are
calculated as concrete examples.}

\begin{document}



\vfill

\eject


\section{Introduction}

Recently, Sen in Ref.\cite{sen} introduced the entropy function
method to calculate the entropy of extremal black holes with near
horizon geometry $AdS_2 \times S^{n-2}$ by defining the entropy of
the extremal black hole to be the extremal limit of the entropy of
a non-extremal black hole so that  one can use the Wald's formula
for entropy given in \cite{wald,mv}. The entropy function method
is an useful approach for computing the entropy from the Wald
formula and it has been generalized to many solutions in
supergravity theory no matter extremal or non-extremal
solutions\cite{ Sen1, Sen2, SSen1, SSen2, SSen3, DSen, Pres, AE1,
AE2, CPTY, Exi, Chand,ag, SS1, SS2, AGJST, CGLP, MS, AGM, CYY,
CDM4, CP1, CP2, CP3, Garousi,caicao}. Actually, the entropy
formula of Wald is based on the first law of black hole mechanics
and it might not work for extremal black holes because of the
vanishing surface gravity and vanishing bifurcation surface. On
the other hand, the entropy function method works for extremal
black holes with near horizon geometry $AdS_2\times S^{n-2}$, and
it might be difficult to apply this method to non-extremal black
holes or more general conditions. This is because there is a
strong hypothesis for the entropy function method \cite{sen
review}: \textit{in any general covariant theory of gravity
coupled to matter fields, the near horizon geometry of a
spherically symmetric extremal black hole in $n$ dimensions has
$SO(2,1)\times SO(D-1)$ isometry}. Apparently, the near horizon
geometry of a non-extremal black hole might not satisfy the above
argument.

The purpose of this paper is to compute the entropy in (anti-)de
Sitter (dS) space with higher derivative gravity terms. Basing on
the method developed by Cai and Cao \cite{caicao}, in this paper,
we will show that a simple integration formula can quickly give
the entropy of neutral black holes in dS and AdS spacetime,
including the higher derivative corrections to the entropy. Black
holes in dS and AdS sapcetimes are extremely important in many
aspects\cite{spkim}. They have special interest from the
holographic point of view due to the well known AdS/CFT
correspondence (dS/CFT correspondence)\cite{mal,strominger}. The
Hawking-Page transition for Schwarzschild-AdS (SAdS) black holes
plays an important role in AdS/CFT correspondence where it was
interpreted by Witten \cite{mal} as the confinement-deconfinement
transition in dual gauge theory. In this sense, Schwarzschild-AdS
is the important tool to describe thermodynamics of CFT and give
crucial support of AdS/CFT correspondence.

Before going to concrete computation, we first summarize the
differences between the method developed by Sen and the method to
be used in this work. In general, the Sen's entropy function
method is
composed of the following three arguments\cite{sen}: \\
1). For an extremal charged black hole with near horizon geometry
$AdS_2\times S^{n-2}$. The part of $AdS_2$ metric is deformed into
$v_1(-r^2dt^2+dr^2/r^2)$, while the $S^{n-2}$ part of the metric
has the form $v_2d\Omega^2_{n-2}$. $v_1$ and $v_2$ are regarded as
constant value here. The deformed metric is assumed as a solution
of the equation of motion for a special action, where gravity is
coupled to a set of electric and magnetic fields and neutral
scalar
fields $u_s$.\\
2). Define an entropy function over the horizon $S^{n-2}$, which
is a function of $v_i$ and $u_s$, and the electric and magnetic
field
($e_i, p_i$).  \\
3).One can find that for given $e_i, p_i$, the values $u_s$ of the
scalar fields as well as the sizes $v_1$ and $v_2$ of $AdS_2$ and
$S^{n-2}$ are determined by extremizing the entropy function with
respect to the variables $u_s$, $v_1$ and $v_2$. The entropy
function at its extremum point is proportional to the entropy of
black holes.

In the following, we can see that for dS and AdS spacetimes:\\
 i). Extremizing the entropy function is not required. And also, we need not deform the metric into the near horizon geometry $AdS_2\times
 S^{n-2}$. The configurations of spacetime can be used directly
 without any change in calculating the entropy. The near horizon limit is also unnecessary in obtaining dS entropies including higher derivative
 gravity terms.\\
 ii). We do not need to assume the radii of dS and AdS space as some constants, such as $v_1$ and
 $v_2$. (One may simply assume these parameters as $1$. Actually $v1$ and $v2$ are quite trivial in calculating
  extremal stringy black hole entropy, because
 in the end their values are always $1$, for example, see \cite{ag,Garousi})\\
 iii). Especially for pure dS and AdS space,  one can still obtain the modified entropy formula without knowing
 the solutions of higher derivative gravity.

The organization of this paper is as follows. In next section we
give a brief review of previous work on Noetherian entropy and
entropy function. Some useful formulae are derived from black hole
thermodynamics. Section 3 discusses the entropy of pure dS space
using our new method. DS entropy with Gauss-Bonnet correction is
calculated in Section 4, where we compare our result with the
entropy obtained by Wald's formula. In section 5 we deal with the
dS entropy in $R^2$ gravity theory. The entropy of black holes in
AdS is investigated in Section 6. The $R^4$ correction to the
entropy is included in this section. The last section contains our
main conclusions and discussions.

\section{Brief review of Noetherian entropy and entropy function}

In this section, we will briefly review some important results
made in the previous work\cite{sen,wald,caicao}, following the
framework of Lagrangian field theories developed by Wald and
viewing the Lagrangian as an $n$-form $\mathbf{L}(\psi)$, where
$\psi=\{g_{ab},R_{abcd},\Phi_{s}, F^I_{ab}\}$ denotes the
dynamical fields considered in this paper, including the spacetime
metric $g_{ab}$, the corresponding Riemann tensor $R_{abcd}$, the
scalar fields $\{\Phi_s$, $s=0,1,\cdots\}$, and the $U(1)$ gauge
fields $F^I_{ab}=\partial_a A_b^I-\partial_b A_a^I$ with the
corresponding potentials $\{A_a^I$, $I=1,\cdots\}$. Under this
definition, the variation of $\mathbf{L}$ is
\begin{equation}
\delta\mathbf{L}=\mathbf{E}_{\psi}\delta \psi +d \mathbf{\Theta},
\end{equation}
where $\mathbf{\Theta}$ is an $(n-1)$-form, which is called {\it
symplectic potential form}, $\mathbf{E}_{\psi}$ corresponds to the
equations of motion for the metric and other fields. Let $\xi$ be
any smooth vector field on the space-time manifold, then one can
define a {\it Noether current form} as
\begin{equation}
\label{Noethercurrent}
\mathbf{J}[\xi]=\mathbf{\Theta}(\psi,\mathcal{L}_{\xi} \psi)-\xi
\cdot \mathbf{L}.
\end{equation}
The fact that $d\mathbf{J}[\xi]=0$ will be preserved when the
equations of motion are satisfied shows that a locally constructed
$(n-2)$-form $\mathbf{Q}[\xi]$ can be introduced and an ``on
shell'' formula can be obtained
\begin{equation}
\mathbf{J}[\xi]=d\mathbf{Q}[\xi]\, \label{onshell}.
\end{equation}
Wald's analysis based on the first law of black hole
thermodynamics showed that for general stationary black holes, the
black hole entropy is a kind of Noether charge at horizon
\cite{wald} and can be expressed as
\begin{equation}
S_{BH}=2\pi \int _{\mathcal{H}} \mathbf{Q}[\xi]\, ,
\end{equation}
where $\xi$ represents the Killing field on the horizon, and
$\mathcal{H}$ is the bifurcation surface of the horizon. It should
be noted that the Killing vector field has been normalized to have
unit surface gravity.

In fact, we can even go further, the definition of the Noether
charge $\mathbf{Q}[\xi]$ can be extended in an arbitrary manner to
$\psi$ which do not satisfy the equations of motion. By contrast,
we call this the so-called ``off shell" form of the Noether
charge, which is defined by the relation~\cite{Iwald}
\begin{equation}
\mathbf{J}[\xi]=d\mathbf{Q}[\xi]+\xi^a\mathbf{C}_a\,
,\label{offshell}
\end{equation}
where $\mathbf{C}_a$ is locally constructed out of the dynamical
fields in a covariant manner and $\mathbf{C}_a=0$ reduces to the
previous definition (\ref{onshell}) ``on shell''. Noether charge
defined in (\ref{offshell}) can be written as
\begin{equation}
\label{conservedcharge}
\mathbf{Q}=\mathbf{Q}^{F}+\mathbf{Q}^g+\cdots
\end{equation}
with
\begin{equation}
\mathbf{Q}^F_{a_1\cdots a_{n-2}}=\frac{\partial
\mathcal{L}}{\partial F^{I}_{ab}}\xi^c A_c^{I}\mbox{{\boldmath
$\epsilon$}}_{aba_1\cdots a_{n-2}}\, ,
\end{equation}
\begin{equation}
\mathbf{Q}^g_{a_1\cdots
a_{n-2}}=-\frac{\partial\mathcal{L}}{\partial
R_{abcd}}\nabla_{[c}\xi_{d]}\mbox{{\boldmath
$\epsilon$}}_{aba_1\cdots a_{n-2}}\, .
\end{equation}
The $``\cdots"$ terms are not important for our following
discussion, so we brutally drop them at first. The relevant
discussion can be found in a recent paper~\cite{caicao}.

On the other hand, A. Sen observed that the entropy of a kind of
extremal black holes which have the near horizon geometry
$AdS_2\times S^{n-2}$ can be obtained by extremizing the so-called
``entropy function'' $\mathbf{f}$ with respect to the moduli on
the horizon\cite{sen}
\begin{equation}\label{efunction}
S_{BH}=2\pi\mathbf{f}=2\pi \left(e_iq_i-f(\vec{u}, \vec{v},
\vec{e},\vec{p})\right).
\end{equation}
where $f$ is defined by
\begin{equation}
\label{ffunction} f(\vec{u}, \vec{v}, \vec{e},\vec{p})=\int
dx^2\cdot\cdot\cdot dx^{n-1}\sqrt{-det g}\mathcal{L}.
\end{equation}
Here $\sqrt{-det g}\mathcal{L}$ is the Lagrangian density,
expressed as a function of the metric $g_{\mu\nu}$, the scalar
fields ${\phi_s}$, the gauge field strength $F^{(i)}_{\mu\nu}$ and
covariant derivatives of these fields.

It was shown in \cite{sen} that if we denote by
$\mathcal{L}_{\lambda}$ a deformation of $\mathcal{L}$ in which we
rescale all factors of Riemann tensor
$R_{\alpha\beta\gamma\delta}$ by $\lambda
R_{\alpha\beta\gamma\delta}$ and define on the near horizon
geometry
\begin{equation}
f_{\lambda}=\sqrt{-det \ g} \mathcal{L}_{\lambda},
\end{equation}
the following relation may be found
\begin{equation}
\left.\frac{\partial f_{\lambda}}{\partial \lambda}
\right|_{\lambda=1}=\int_{\mathcal{H}}\sqrt{-det \ g}
R_{\alpha\beta\gamma\delta}\frac{\partial\mathcal{L}}{\partial
R_{\alpha\beta\gamma\delta}}dx^{1}\cdots
dx^{n-2}=f-e_i\frac{\partial f}{\partial e_i},
\end{equation}
where $\alpha, \beta, \gamma, \delta$ are summed over the
coordinates $r$ and $t$.

Now following our previous work on black hole thermodynamics and
entropy function \cite{gs} we give an alternative way to calculate
the black hole entropy. We consider an $n$--dimensional
spherically symmetric black hole with the metric in the form of
\begin{equation}
ds^2=-a^2(r)
dt^2+\frac{dr^2}{a^2(r)}+b^2(r)d\Omega_{n-2}^2,\label{metric}
\end{equation}
where $a$ and $b$ are functions of $r$, and $d\Omega_{n-2}^2$ is
the line element for $ {S}^{n-2}$.  Since $\mathbf{\Theta}=0$ if
$\xi$ is a Killing vector, we find by integrating over a Cauchy
surface $\bf{\mathcal{C}}$ on Eq.(\ref{onshell})
\begin{equation}
\label{interJ}
\int_{\bf{\mathcal{C}}}\mathbf{J}=-\int_{\bf{\mathcal{C}}}\xi
\cdot
\mathbf{L}=\int_{\bf{\mathcal{C}}}d\mathbf{Q}[\xi]=\int_{\infty}\mathbf{Q}-\int_{\mathcal{H}}\mathbf{Q}\,
\end{equation}
where $\mathcal{H}$ denotes the interior boundary, and we have
used the Stokes theorem.  For an asymptotically flat, static
spherically symmetric black hole, one can simply choose
$\xi=\partial_t=\frac{\partial}{
\partial t}$, then the free energy of the system is shown to be
\cite{gs}
\begin{eqnarray}
\label{conserve}
F=\mathcal{E}-\int_{\mathcal{H}}\mathbf{Q}[\xi^{a}],
\end{eqnarray}
where $F=TI_E$ with $T$ and $I_E$ the temperature and Euclidean
action respectively. $\mathcal{E}$ in above formula is the
``canonical energy '', which is defined by\cite{wald},
\begin{eqnarray}\label{energy}
\mathcal{E}=\int_{\infty}(\mathbf{Q}[t]-t\cdot \mathbf{B}),
\end{eqnarray}
where $\mathbf{B}$ is an $(n-1)$-form given by
$$
\delta \int_{\infty}t\cdot\mathbf{B}=\int_{\infty}t\cdot
\mathbf{\Theta}.
$$
The variation of Eq.(\ref{conserve}) leads to
\begin{eqnarray}
\label{conserveE} \delta F=\delta
\mathcal{E}-\delta\int_{\mathcal{H}}\mathbf{Q}[\xi^{a}]
\end{eqnarray}
Now by noting that the Noetherian charge can be decomposed into
two parts as shown in (\ref{conservedcharge}), we consider a
stretched region near the horizon ranged from $r_{H}$ to
$r_{H}+\delta r$,
\begin{equation}
\delta\int_{\mathcal{H}}\mathbf{Q}[\xi]=\int_{r_H}^{r_{H}+\delta
 r}\left(\mathbf{Q}^{\rm F}[\xi]+\mathbf{Q}^{g}[\xi]\right)
\end{equation}
The Killing equation gives $\nabla_{[a}\xi_{b]}=2\kappa
\mbox{{\boldmath $\epsilon$}}_{ab}$ (where $\kappa$ is the surface
gravity of the hole), and the two parts are found to be\cite{cai}
\begin{eqnarray}
\label{equation1} &&\int_{r_H}^{r_{H}+\delta
 r}\mathbf{Q}^g[\partial_t] =\delta r\left[\kappa'E+\kappa
E'\right]_{r_H}+\mathcal{O}(\delta
r^2)\, ,\\
\label{equation2} &&\int_{r_H}^{r_{H}+\delta
 r}\mathbf{Q}^F[\partial_t] =q_I e_I\delta r +e_I q_I' \delta r+{\cal O}(\delta r^2).
\end{eqnarray}
where $E(r)$ is defined as
\begin{equation}
E(r) \equiv -\int_{\mathcal{H}}\frac{\partial\mathcal{L}}{\partial
R_{abcd}}\mbox{{\boldmath $\epsilon$}}_{ab}\mbox{{\boldmath
$\epsilon$}}_{cd}dx^{1}\cdots dx^{n-2}\, .
\end{equation}
The above formula is exactly the Wald formula for entropy without
the factor $2\pi$ \cite{wald}.Therefore, we find $E(r)$ is related
to the entropy  by $2\pi E(r_H)=S$. $e_I$ and the $U(1)$
electrical-like charges in Eq. (\ref{equation2}) are defined to be
\begin{eqnarray}
e_I &\equiv& F_{rt}^I(r_H), \label{eq34}\\
q_I &\equiv & \int_{r}\frac{1}{(n-2)!}\frac{\partial
\mathcal{L}}{\partial F^{I}_{ab}}\mbox{{\boldmath
$\epsilon$}}_{aba_1\cdots a_{n-2}}dx^{a_1}\wedge\cdots \wedge
dx^{a_{n-2}}\, \nonumber\\
&=&\frac{\partial }{\partial
e_I}\int_{r_H}\frac{\mathcal{L}}{2(n-2)!}\mbox{{\boldmath
$\epsilon$}}^{ab}\mbox{{\boldmath $\epsilon$}}_{aba_1\cdots
a_{n-2}}dx^{a_1}\wedge\cdots \wedge dx^{a_{n-2}}=\frac{\partial
f(r_H)}{\partial e_I}\, ,
\end{eqnarray}
where we have written $F_{ab}^I(r_H)$ as $e_I \mbox{{\boldmath
$\epsilon$}}_{ab}$. If the near horizon extension
$r_{H}\rightarrow r_{H}+\delta r$ is also done for the free
energy, we find that
\begin{equation}
\label{equation3} \delta F=-\int_{r_{H}}^{r_{H}+\delta r}f
dr=-f(r_{H})\delta r+{\cal O}(\delta r^2)
\end{equation}
Substituting Eqs. (\ref{equation1}), (\ref{equation2}) and
(\ref{equation3}) into Eq. (\ref{conserveE}), we obtain
\begin{equation}
\label{equation4}\mathbf{f}\delta r=-S\delta T,
\end{equation}
where $\mathbf{f}=\left(-f(r_{H})+q_{I}e_{I}\right)$ is the entropy
function for extremal black hole as shown in (\ref{efunction}), and
we have used the relation $\delta\mathcal{E}=T\delta
S-e_{I}q_{I}'\delta r$\footnote{More general form of the first law
of black hole thermodynamics is $\delta\mathcal{E}=T\delta S+
\Phi_I\delta Q_I$, where
$\Phi_I\equiv-\xi^{a}A_a^I\mid_{\mathcal{H}}$ is the electrostatic
potential and $Q_I \equiv \int_{r}\frac{1}{(n-2)!}\frac{\partial
\mathcal{L}}{\partial F^{I}_{ab}}\mbox{{\boldmath
$\epsilon$}}_{aba_1\cdots a_{n-2}}dx^{a_1}\wedge\cdots \wedge
dx^{a_{n-2}}$ is charge. It is not difficult to find that in our
case, $\Phi_I=-e_I$ and $Q_I=q_I$. The standard first law hence can
be recovered by using these relations.}. In the limit $\delta
r\rightarrow 0$, we obtain an equation which governs the entropy
function for non-extremal black holes
\begin{eqnarray}\label{nonextremal}
S T'=-\mathbf{f},
\end{eqnarray}
where prime denotes derivative with respect to $r$.

What we discussed above is the asymptotically flat case. The
non-asymptotically flat case, such as asymptotically dS or AdS
cases, however, is proved to be a little different from the one
discussed above due to the definition of the Hamiltonian. By showing
the difference we will start with looking at the Noetherian
definition of mass in AdS spacetime. The definition of Hamiltonian
is shown to be \cite{dutta}
\begin{equation}\label{ads hamil}
\delta H= \int_{\hat{R}} (\delta {\bf Q}[\xi]-\xi\cdot {\bf
B})-\int_{\hat{R}} \delta ({\bf
Q}_{AdS}[\tilde{\xi}]-\tilde{\xi}\cdot {\bf B}_{AdS}),
\end{equation}
where $\hat{R}$ is a cutoff at an outer boundary. This is unlike the
flat case where the integration occurs at infinity. The second term
in (\ref{ads hamil}) corresponds to a constant by which we preserve
the fact that the energy is zero in pure AdS. Notice that $\xi^a$ is
fixed during the variation, the spacetime and AdS background should
have the same boundary geometry at $r=\hat{R}$, which leads to
$|\tilde{\xi}|^2=|\xi|^2$ on the boundary. In the end, taking into
account the linear relation between $Q[\xi]$ and $\xi$ one obtains
the mass of the system \cite{dutta}
\begin{equation}\label{ads mass}
\mathcal{E}= \int_{\hat{R}} ({\bf Q}[t]-t\cdot {\bf
B})-\left[\left(\frac{g_{tt}}{g_{tt}^{AdS}}\right)^{1/2}\right]_{r=\hat{R}}\int_{\hat{R}}
({\bf Q}_{AdS}[t]-t\cdot {\bf B}_{AdS}),
\end{equation}
where we have taken $\xi$ to be the time translation Killing vector
since we only consider the static case.

We know that $\xi$ vanishes on the bifurcate horizon hence ${\bf
\Theta}=0$ and the Noether current simplifies to $ {\bf J} = d{\bf
Q}= -\xi\cdot {\bf L}$. Integrating it over a Cauchy surface ${\cal
C}$ with the interior boundary {$\cal H$} as the event horizon of
the black hole, and the outer boundary at $r=\hat{R}$, we get \be
\int_{{\cal H}}{\bf Q}[t]={\cal E} +\int_{\cal C}\xi^t\cdot {\bf
L}(g_{BH})+\int_{R}t\cdot {\bf B}+\left[ \left\{\frac{g^{BH}_{tt} }{
g^{AdS}_{tt}}\right\}^{1/2} \right]_{r=\hat{R}} \int_{\hat{R}}
 ({\bf Q}_{AdS}[t]-t\cdot {\bf B}_{AdS}).
\ee On the other hand, one integrates over another Cauchy surface
$\mathcal{C}_{AdS}$ to get

\be \int_{\hat{R}}
  {\bf Q}_{AdS}[t]=-\int_{\mathcal{C}_{AdS}}\xi^t\cdot {\bf L}(g_{AdS}). \ee
Following the procedures made in last section, we variate above
formula, and note
\begin{equation}\delta\int_{{\cal H}}{\bf
Q}[t]=\int_{r_{H}+\delta r}{\bf Q}[t]-\int_{r_{H}}{\bf
Q}[t]=[(TS)'+e_Iq_I+e_I q'_I]\delta r,\end{equation}
and
\begin{eqnarray}
\delta\int_{\mathcal{C}}(\xi^t\cdot {\bf L}+t\cdot{\bf B})=-\delta
F=\int_{r_H}^{r_H+ \delta r}f dr=f(r_{H}) \delta r,
\end{eqnarray}
in the end we obtain our final result \be\label{ads master}
\mathcal{E}'+e_Iq'_I-(TS)' ={\bf f}_{BH}-\left [\left(
    {g^{BH}_{tt} \over g^{AdS}_{tt}}
\right)^{1/2} \right]_{r=\hat{R}}{\bf f}_{AdS},\ee where ${\bf
f}_{BH}=e_Iq_I-f_{BH}$ and ${\bf f}_{AdS}=-f_{AdS}$ denote the
entropy functions of black hole and AdS respectively. To obtain the
correct entropy we should integrate the right-hand side of Eq.
(\ref{ads master}) with appropriate integration region. A reasonable
integration region is shown to be\cite{dutta} \be\label{ads
integration} F=\int_{r_H}^{\hat{R}}{\bf f}_{BH}-\left [\lb
    {g^{BH}_{tt} \over g^{AdS}_{tt}}
\rb^{1/2} \right]_{r=\hat{R}}\int_{0}^{R}{\bf f}_{AdS},\ee where
$F=\mathcal{E}-TS+e_Iq'_I$ is the free energy of the system. The
entropy thus can be obtained by
$$S=-\frac{\partial F}{\partial T}.$$
In particular, the corresponding expression for pure dS (or AdS)
is
\begin{equation}
\label{ds} TS=\int_{0}^{r_{H}}{\bf f}_{dS}dr,
 \end{equation}
where $r_H$ is the event horizon of dS space. We can show the
entropy of dS space in any $R^2$ gravity theories can be computed
by\footnote{The details can be found in Appendix \ref{appendixA}.}
\begin{eqnarray}\label{ds correction}
S=S_0+\gamma S_1=\frac1{T_{0}}\int_0^{r_H^{(0)}}\left({\bf
f}_0+\gamma{\bf f}_1^{(0)}\right)dr,
\end{eqnarray}
where superscript ``$(0)$'' denotes the variables computed by using
the unperturbative metric, and $\mathbf{f}_0$ and $\mathbf{f}_1$
(including $\mathbf{f}_1^{(0)}$ and $\mathbf{f}_1^{(1)}$, see
Appendix \ref{appendixA} for detail) represent entropy function with
and without higher derivative corrections respectively. Both
$\mathbf{f}_0$ and $T_{0}$ are calculated by using the
unperturbative metric. $\gamma$ is a small quanta showing the
coupling strength. This implies one can obtain the entropy of dS
spacetime in any $R^2$ gravity theories without knowing the
corrected metric. In our present paper, we will confirm this
argument by taking several examples.

\section{Entropy of $D$-dimensional de Sitter spacetime}

In this section, we will give a simple example to demonstrate how
our method works for non-supergavity spacetime. We may calculate
entropy of pure dS spacetime by using the method discussed in the
last section.

From the traditional point of view, the ``entropy function''
method may break down when we apply it to calculating the entropy
of dS  spacetime due to its non-extremality. Our analysis in the
last section indicates we may end this embarrassed situation of
this kind of space by extending the entropy function to the case
including non-extremal spacetime. The main idea is to relate the
entropy function with the black hole thermodynamics with the help
of the entropy definition of Wald, whose definition is valid for
any non-extremal black holes. By doing so, we obtained a formula
by which one can compute the dS entropy, \textit{i.e.},
\begin{equation}\label{ds2}
TS=\int_{0}^{l}{\bf f}_{dS}dr,
 \end{equation}
where $l$ denotes the cosmological radius. The Einstein-Hilbert
action with a positive cosmological constant is given by
\begin{equation}\label{action}
I= \frac1{16\pi G_D} \int d^{D}x \sqrt{-det\ g} ( R-2\Lambda),
\end{equation}
where $R$ is the Ricci scalar of the spacetime manifold,
$\Lambda=(D-1)(D-2)/2l^2$ is the cosmological constant, and $G_{D}$
is the $D$-dimensional Newton constant. The corresponding static
metric is given by
\begin{equation}\label{metric}
ds^2=-\left(1-\frac{r^2}{l^2}\right)dt^2+\left(1-\frac{r^2}{l^2}\right)^{-1}dr^2+r^2
d\Omega_{D-2}^2.
\end{equation}
Recalling the definition of $f$ in (\ref{ffunction}), we obtain
\begin{equation}\label{entropyfunction}
\mathbf{f}_{dS}=-\frac{(D-1)\pi^{\frac{D-3}{2}}r^{D-2}}{4 G_D
\Gamma(\frac{D-1}2)l^2}.
\end{equation}
The temperature for $D$-dimensional dS spacetime is
\begin{equation}
T=-\frac{1}{2\pi l},
\end{equation}
Direct calculation of Eq. (\ref{ds2}) gives the entropy of dS
spacetime
\begin{equation}\label{dsentropy}
 S_{dS}=\frac{A}{4G_D}=\frac{l^{D-2}A_{D-2}}{4G_D},
\end{equation}
where $A_{D-2}=2\pi^{(D-1)/2}/\Gamma((D-1)/2)$ is the area of the
$(D-2)$ dimensional unit sphere. (\ref{dsentropy}) is exactly the
Bekenstein-Hawking entropy of dS spacetime. This indicates that
our method discussed in Sec. 2 works well for dS space.


\section{Entropy with Gauss-Bonnet term in de Sitter space}

In this section, we wish to confirm our argument made in Sec.2,
\textit{i.e.}, \emph{one can obtain the entropy of dS spacetime in
any $R^2$ gravity theories without knowing the corrected metric}. In
particular, we shall consider a specific higher derivative
correction to the action---the Gauss-Bonnet term. This term, which
is generated from the heterotic and bosonic string theory low energy
effective theory, is a natural correction term to the
Einstein-Hilbert action (\ref{action}). It corresponds to an
additional term in the Lagrangian density of the form
\cite{zwiebach}\footnote{What we should mention is that we require
$D\geq 5$ in this case since the Gauss-Bonnet term is a topological
invariant in four dimension.}
\begin{equation}
\mathcal{L}_{GB}=\frac{\alpha}{16\pi
G_D}\left\{R_{\mu\nu\rho\sigma}R^{\mu\nu\rho\sigma}-4R_{\mu\nu}R^{\mu\nu}+R^2\right\},
\end{equation}
where $\alpha$ is the coupling constant with dimensions
$(length)^2$. In particular, $\alpha=\alpha'/4$ in the low energy
effective action of heterotic string theory. Then the action
containing the Gauss-Bonnet term becomes \cite{deser}
\begin{equation}\label{actiongb}
I= \frac1{16\pi G_D} \int d^{D}x \sqrt{-det\ g} \left(
R-2\Lambda+\mathcal{L}_{GB}\right).
\end{equation}
The corresponding equation of motion is\cite{cai}
\begin{eqnarray}\label{eom}
&&R_{\mu\nu}-\frac12 R
g_{\mu\nu}+\frac{(D-1)(D-2)}{2l^2}g_{\mu\nu}=\nonumber\\&&\alpha
\left\{\frac12 g_{\mu\nu}
\mathcal{L}_{GB}-2\left(R_{\mu\alpha\beta\gamma}R_{\nu}
^{\alpha\beta\gamma}-2R^{\alpha\beta}R_{\mu\alpha\nu\beta}-2R_{\mu}
^{\alpha}R_{\nu\alpha}+R R_{\mu\nu}\right)\right\}
\end{eqnarray}


\subsection{Entropy function method}

We wish to find the leading order correction in $\alpha$ to the
entropy of the dS spacetime. Usually, before we do this, we should
first obtain the corresponding solutions to the modified Einstein
equation (\ref{eom}). This, however, turns out to be not so easy
except for some special cases. Fortunately, the entropy function
method provides us with an elegant way to calculate the entropy of
higher derivative gravity without knowing the corrected metric.
The Gauss-Bonnet term corresponding to the metric (\ref{metric})
is then given by
\begin{equation}\label{gb}
\mathcal{L}_{GB}=\frac{D(D-1)(D-2)(D-3)\alpha}{16\pi G_D l^4}.
\end{equation}
The definition of $f$ in (\ref{ffunction}) shows Lagrangian
density with higher derivative correction gives unavoidably an
extra term of $f$. In Gauss-Bonnet Einstein gravity, this term
comes from the Gauss-Bonnet term as shown in (\ref{gb}). This in
turn will change the expression of entropy function as defined in
previous section. Direct computation shows the correction term to
the entropy function (\ref{entropyfunction}) is
\begin{eqnarray}
\nonumber \mathbf{f}_{1GB}&=&-f_{GB}=-\int d^{D-2}x \sqrt{-det g}\mathcal{L}_{GB}\\
&=&-\frac{\alpha D(D-1)(D-2)(D-3)A_{D-2}}{16\pi G_D l^4}r^{D-2}.
\end{eqnarray}
Substituting this into entropy function equation (\ref{ds
correction}), we can easily obtain the entropy generated by
Gauss-Bonnet correction term
\begin{equation}
S_{1GB}=\frac{\int_0^{l} \mathbf{f}_{1GB}dr}{T}=\frac{\alpha
D(D-2)(D-3)A_{D-2}}{8 G_D}l^{D-4}.
\end{equation}
Hereafter we do not distinguish $T$ from $T_{0}$. Consequently,
the entropy of dS space (near the horizon) including Gauss-Bonnet
correction becomes
\begin{equation}\label{gbentropy}
S=S_{dS}+ S_{1GB}=\frac{l^{D-2}A_{D-2}}{4G_D}\left(1+\frac{\alpha
D(D-2)(D-3)}{2l^2}\right)+\mathcal{O}(\alpha^2).
\end{equation}


\subsection{Wald's approach}

In this subsection, we wish to check our result (\ref{gbentropy})
obtained by entropy function method. One way is to use the Wald's
approach. To do this, we have to find the corrected metric of the
modified equation of motion (\ref{eom}) before we compute the
corrected entropy of the Gauss-Bonnet gravity. Generally speaking,
it is not easy to obtain the solution of this equation of motion.
However, in some special cases, we can find the exact solution.
Now we assume the metric to be of the form as discussed in
\cite{deser}
\begin{equation}
ds^2=-e^{2\nu} dt^2+e^{2\lambda} dr^2+r^2 d \Omega^2_{D-2},
\end{equation}
where $\nu(r)$ and $\lambda(r)$ are functions of $r$ only. The
solution under this assumption is shown to be\cite{deser}
\begin{equation}\label{metricgb}
e^{2\nu}=e^{-2\lambda}=1+\frac{r^2}{2\tilde{\alpha}}\left(1\mp\sqrt{1+\frac{4\tilde{\alpha}}{l^2}}\right),
\end{equation}
where $\tilde{\alpha}=(D-3)(D-4)\alpha$. As shown in
\cite{deser,cai}, the branch with the ``$+$'' sign is unstable and
the graviton is a ghost. Therefore we only consider the case with
``$-$'' sign in this paper. Consequently, the metric becomes
\begin{equation}\label{metricgb}
e^{2\nu}=e^{-2\lambda}=1+\frac{r^2}{2\tilde{\alpha}}\left(1-\sqrt{1+\frac{4\tilde{\alpha}}{l^2}}\right).
\end{equation}
Since the horizon of this geometry occurs where $e^{2\nu}=0$, we
find the event horizon radius
\begin{eqnarray}\label{horizongb}
\nonumber
r_H&=&\left(\frac{2\tilde{\alpha}}{\sqrt{1+4\tilde{\alpha}/l^2}-1}\right)^{1/2}\\
&=& l+\frac{\tilde{\alpha}}{2l}+\mathcal{O}(\tilde{\alpha}^2).
\end{eqnarray}

One may expect that we can obtain the Bekenstein-Hawking entropy
by direct substituting (\ref{horizongb}) into the entropy-area
formula $ S_{BH} = \frac A4$. This, however, has been proved to be
no longer true by a lot of earlier investigations\cite{myers}. It
was shown in \cite{jacobson} that $S$ should take the form of a
geometric expression evaluated at the event horizon. There are
many ways to calculate the entropy of this spacetime: one way is
to use the Wald's formula \cite{myers1}; the other is related to
the Euclidean entropy of the space as shown in \cite{dutta,cho}.
It also can be obtained by assuming the spacetime satisfies the
first law of thermodynamics as shown in \cite{ross}. As an
example, we use Wald's approach to check our method obtained in
the last subsection. Using the Wald's formula, we know that the
entropy of a hole valid to any effective gravitational action
including higher curvature interactions is given by\cite{myers1}
\begin{equation}
S_{Wald}=\frac{1}{4G_D}\int_{\mathcal{H}}d^{D-2}x\sqrt{h}\left[1+2\alpha
\tilde{R}(h)\right],
\end{equation}
where $h_{ij}$ is the induced metric on the horizon,\footnote{In
present case, $h_{ij}$ denotes the metric of sphere $S^{D-2}$.} and
$\tilde{R}(h)=h^{ij}h^{kl}R_{ikjl}$ is the Ricci scalar calculated
by using the induced metric $h_{ij}$. The entropy calculated in this
way is proved to be \cite{cho}
\begin{eqnarray}
\nonumber S&=&\frac{r_H^{D-2}A_{D-2}}{4G_D}\left(1+\frac{2(D-2)\tilde{\alpha}}{(D-4)r_H^2}\right)\\
&=&\nonumber
\frac{l^{D-2}A_{D-2}}{4G_D}\left(1+\frac{D(D-2)\tilde{\alpha}}{2(D-4)l^2}\right)+\mathcal{O}(\tilde{\alpha}^2)\\
&=&
\frac{l^{D-2}A_{D-2}}{4G_D}\left(1+\frac{D(D-2)(D-3)\alpha}{2l^2}\right)+\mathcal{O}(\alpha^2),
\end{eqnarray}
which is exactly the same as entropy (\ref{gbentropy}) obtained by
the entropy function method. Therefore we conclude that, compared
with other methods (such as the entropy function method
proposed by Sen and the Wald's approach), our method has many advantages. Roughly speaking, they are:\\
 i). Unlike the standard entropy function method, we do not need to extremize the entropy function, and also, we need not deform the metric into the near horizon geometry.
 The configurations of spacetime can be used directly
 without any change in calculating the entropy. The near horizon limit is also unnecessary.\\
 ii).Unlike  the standard entropy function method, we do not need to introduce some constants into the metric, such as $v_1$ and
 $v_2$.\\
 iii).Unlike the Wald's approach or Euclidean approach\cite{dutta}, we do not need to know the corrected metric of higher derivative gravity for dS space.


\section{Entropy with $R^2$ term in de Sitter space}

In this section, we consider $R^2$ term which involves a
correction proportional to
$R_{\mu\nu\alpha\beta}R^{\mu\nu\alpha\beta}$. This term, which is
the first term that one can add to the Einstein-Hilbert action,
has been discussed in several previous works \cite{dutta,cho}. It
corresponds to an additional term in the Lagrangian density of the
form \cite{zwiebach}
\begin{equation}
\mathcal{L}_{R^2}=\frac{\alpha}{16\pi
G_D}R_{\mu\nu\rho\sigma}R^{\mu\nu\rho\sigma},
\end{equation}
where again $\alpha$ is the coupling constant with dimensions
$(length)^2$, and $\alpha=\alpha'/4$ in the low energy effective
action of heterotic string theory. The action containing this term
becomes
\begin{equation}\label{actionr2}
I= \frac1{16\pi G_D} \int d^{D}x \sqrt{-det\ g} \left(
R-2\Lambda+\mathcal{L}_{R^2}\right).
\end{equation}
The corresponding equation of motion is\cite{dutta}
\begin{eqnarray}\label{eomr2}
\nonumber R_{\mu\nu}-\frac{R g_{\mu\nu}}2
+\frac{(D-1)(D-2)}{2l^2}g_{\mu\nu}&&= \frac{\alpha}2 g_{\mu\nu}
\mathcal{L}_{R^2}-2\alpha\left(R_{\mu\alpha\beta\gamma}R_{\nu}
^{\alpha\beta\gamma}+2R^{\alpha\beta}R_{\mu\alpha\nu\beta}\right.\\&&\left.-2R_{\mu}
^{\alpha}R_{\nu\alpha}+2\Box
R_{\mu\nu}-\frac12(\nabla_{\mu}\nabla_{\nu}+\nabla_{\nu}\nabla_{\mu})R
\right).
\end{eqnarray}


\subsection{Entropy function method}

In this subsection, we evaluate the $R^2$ correction to the
entropy of the dS spacetime. As mentioned in the last section, we
do not need to know the modified metric before compute the
corrected entropy by using the entropy function method. Therefore
we start with calculating the $R^2$ term corresponding to the
uncorrected metric (\ref{metric})
\begin{equation}\label{r2}
\mathcal{L}_{R^2}=\frac{D(D-1)\alpha}{8\pi G_D l^4}.
\end{equation}
The definition of $f$ in (\ref{ffunction}) shows Lagrangian
density with higher derivative correction gives unavoidably an
extra term of $f$ which comes from the $R^2$ term as shown in
(\ref{r2}). This in turn will change the expression of entropy
function as defined in previous section. Direct computation shows
the correction term to the entropy function
(\ref{entropyfunction}) is
\begin{equation}
\mathbf{f}_{1R^2}=-\frac{\alpha D(D-1)A_{D-2}}{8\pi G_D
l^4}r^{D-2}.
\end{equation}
Substituting this into Eq. (\ref{ds correction}), we can easily
obtain the entropy function generated by $R^2$ correction term
\begin{equation}
S_{1R^2}=\frac{\alpha DA_{D-2}}{4 G_D}l^{D-4}.
\end{equation}
Consequently, the entropy of dS space including $R^2$ correction
becomes
\begin{equation}\label{r2entropy}
S=S_{dS}+ S_{1R^2}=\frac{l^{D-2}A_{D-2}}{4G_D}\left(1+\frac{\alpha
D}{l^2}\right)+\mathcal{O}(\alpha^2).
\end{equation}


\subsection{Wald's approach}

In last subsection, we have calculated the corrected entropy for
dS space by using entropy function method. To confirm our result,
we now appeal to Wald's approach. To do this, we should first find
the corrected metric of the modified equation of motion
(\ref{eomr2}). Generally speaking, it is not easy to obtain the
solution of this equation of motion. However, in some special
case, we can find a perturbative metric solution. Now we assume
the metric to be of the form
\begin{equation}
ds^2=-e^{2\nu} dt^2+e^{2\lambda} dr^2+r^2 d \Omega^2_{D-2},
\end{equation}
where $\nu$ and $\lambda$ are functions of $r$ only, and have the
form
\begin{eqnarray}\label{horizonr2}
\nonumber
e^{2\nu}&=& e^{2\nu_0} (1+\alpha \epsilon(r)), \\
e^{2\lambda}&=& e^{-2\nu_0} (1+\alpha \mu(r)),
\end{eqnarray}
where $e^{2\nu_0}=1-r^2/l^2$, $\epsilon(r)$ and $\mu(r)$ are some
undetermined functions of $r$, which can be obtained by evaluating
the equation of motion (\ref{eomr2}) perturbatively. The result
reads
\begin{equation}\label{metricr2}
\epsilon(r)=-\mu(r)=e^{-2\nu_0}\frac{2(D-4)r^2}{(D-2)l^4}.
\end{equation}
Consequently, the metric becomes
\begin{equation}\label{metricr2}
e^{2\nu}=e^{-2\lambda}=1-\frac{r^2}{l^2}+\frac{2\alpha(D-4)r^2}{(D-2)l^4},
\end{equation}
Since the horizon of this geometry occurs where $e^{2\nu}=0$, we
see
\begin{eqnarray}\label{horizonr2}
\nonumber
r_H&=&\frac{l}{\sqrt{1-\frac{2(D-4)\alpha}{(D-2)l^2}}}\\
&=& l+\frac{(D-4)\alpha}{(D-2)l}+\mathcal{O}(\alpha^2).
\end{eqnarray}
Using the Wald's formula, we know that the entropy of a hole valid
to any effective gravitational action including higher curvature
interactions is given by \cite{myers1}
\begin{equation}
S_{Wald}=\frac{1}{4G_D}\int_{\mathcal{H}}d^{D-2}x\sqrt{h}\left[1+2\alpha
\left(R-2h^{ij}R_{ij}+\tilde{R}(h)\right)\right],
\end{equation}
where again $\tilde{R}(h)=h^{ij}h^{jk}R_{ikjl}$. Direct
computation shows
\begin{eqnarray}
\tilde{R} &=& \frac{(D-2)(D-3)}{r_H^2}+\mathcal{O}(\alpha)\\
h^{ij}R_{ij}&=& \frac{(D-1)(D-2)}{l^2}+\mathcal{O}(\alpha)\\
R&=& \frac{D(D-1)}{l^2}+\mathcal{O}(\alpha).
\end{eqnarray}
Therefore, the entropy calculated in this way is
\begin{eqnarray}
\nonumber S&=&\frac{r_H^{D-2}A_{D-2}}{4G_D}\left[1+2\alpha\left\{\frac{(D-2)(D-3)}{r_H^2}-\frac{(D-1)(D-4)}{l^2}\right\}\right]\\
&=&\nonumber
\frac{l^{D-2}A_{D-2}}{4G_D}\left[1+\frac{(D-4)\alpha}{l^2}\right]\left(1+\frac{4\alpha}{l^2}\right)+\mathcal{O}(\alpha^2)\\
&=&
\frac{l^{D-2}A_{D-2}}{4G_D}\left(1+\frac{D\alpha}{l^2}\right)+\mathcal{O}(\alpha^2),
\end{eqnarray}
which is exactly the same as entropy (\ref{r2entropy}) obtained by
the entropy function method.

\section{Black hole Entropy in $5-$dimensional (anti-) de Sitter space}

In previous sections, we investigate the entropy function for pure
dS space by using a powerful method developed in \cite{gs}. We
showed that entropy obtained by this method agrees with the one
computed by other approaches (say, Wald's approach) very well.
Moreover, for that case we can obtain the correct entropy with
higher derivative corrections without knowing the modified metric.
A question is that what will happen if there is a black hole in
the dS (or AdS) background. How to calculate the entropy of this
case especially when there are higher derivative corrections? In
this section, we try to answer this question by applying our
method to a $5$-dimensional black hole in the Type IIB string
theory. We will show that, although it is no longer possible to
obtain the entropy without knowing the modified metric when higher
curvature corrections are included, our method is still applicable
to this case.

\subsection{Entropy of Schwarzschild-AdS and Schwarzschild-dS black hole}

The $5-$dimensional IIB superstring effective action which is
obtained by compactifying the ten dimensional action on the $S^5$
is \cite{gubser}\footnote{We have set the dilaton to a constant
and we omit it in this paper.}
\begin{eqnarray}\label{iibaction}
I &=& \frac{1}{16\pi G_{5}}\int d^{5}x \sqrt{-det\ g} \left[R-2
\Lambda\right].
\end{eqnarray}

We first pay our attention to the Schwarzschild-AdS black hole. In
the throat approximation, $r\ll l$, a solution corresponding to
the AdS black hole is \cite{gubser,maldacena}
\begin{eqnarray}\label{5dmetric}
ds^2=\frac{r^2}{l_{AdS}^2}\left[-\left(1-\frac{r_0^4}{r^4}\right)dt^2+
d\vec{x}^2\right]+\frac{l_{AdS}^2}{r^2}\left(1-\frac{r_0^4}{r^4}\right)^{-1}dr^2
\end{eqnarray}
The Lagrangian corresponding to this metric reads
\begin{equation}\label{adsgb}
\mathcal{L}=-\frac{1}{2 \pi G_5 l_{AdS}^2}.
\end{equation}
Therefore the entropy function can be obtained by its definition
\begin{eqnarray}
\mathbf{f}_{BH}=-f_{BH}=-\int d^3\vec{x}\sqrt{-det\ g}\mathcal{L}=
\frac{V_3 r^3}{2 \pi G_5 l_{AdS}^5},
\end{eqnarray}
where $V_3=\int d^3 \vec{x}$ is the volume. On the other hand, the
entropy function for pure AdS background is given by
\begin{eqnarray}
\mathbf{f}_{AdS}=-\int d^3\vec{x}\sqrt{-det\
g_{AdS}}\mathcal{L}_{AdS}= \frac{V_3 r^3}{2 \pi G_5 l_{AdS}^5},
\end{eqnarray}
which is the same as the one of the black hole. Therefore from Eq.
(\ref{ads integration}) we obtain the free energy of the system
(Note that in present case, $r_+=r_0$)
\begin{eqnarray}
F&=&\int_{r_0}^{R}{\bf f}_{BH}-\left [\left(
    {g^{BH}_{tt} \over g^{AdS}_{tt}}
\right)^{1/2} \right]_{r=R}\int_{0}^{R}{\bf f}_{AdS},\\
&\simeq& -\frac{V_3 r_0^4}{16 \pi G_5 l_{AdS}^5}.
\end{eqnarray}
According to the definition (\ref{ads mass}), the ADM energy in this
case is given by \cite{dutta}
\begin{eqnarray}
\mathcal{E}&=&\int_R \mathbf{Q}[t]-\int_R
\mathbf{Q}_{AdS}[\tilde{t}]\\
&=&\int_R dS_{ab}\sqrt{- det\ g}Q^{ab}-\left [\left(
    {g^{BH}_{tt} \over g^{AdS}_{tt}}
\right)^{1/2} \right]_{r=R}\int_R dS_{ab}\sqrt{- det\ g_{AdS}}
Q_{AdS}^{ab},
\end{eqnarray}
where
$$
Q^{ab}=\frac{\kappa}{8\pi G_5},
$$
and $\kappa=\left(\frac{r}{l^2}+\frac{r_0^4}{l^2r^3}\right)\big
|_{r=r_0}$ is the surface gravity of the black hole. In the end we
obtain
\begin{eqnarray}
\mathcal{E}=\frac{3V_3 r_0^4}{16\pi G_5l_{AdS}^5}.
\end{eqnarray}
The temperature at the horizon for this black hole AdS spacetime
is
\begin{equation}\label{sg}
T=\frac{r_0}{\pi l_{AdS}^2},
\end{equation}
Therefore using the relation
\begin{equation}
TS=\mathcal{E}-F,
\end{equation}
we obtain the entropy of Schwarzschild-AdS black hole
\begin{equation}\label{adsentropy}
S=\frac{V_3 r_0^3}{4 G_5 l_{AdS}^3}.
\end{equation}
This is exactly the Bekenstein-Hawking entropy for this black
hole\cite{dutta}.

It is not difficult to check that following the above procedure,
we can obtain the entropy of the Schwarzschild-dS black hole. The
result is proved to be the same form as the one for the
Schwarzschild-AdS black hole, \emph{i.e.},
\begin{equation}
S=\frac{V_3 r_0^3}{4 G_5 l_{dS}^3}.
\end{equation}


\subsection{$R^4$ correction to the entropy}

In this subsection, due to the AdS/CFT (or dS/CFT) correspondence,
we wish to study the corrections that appear in the Type IIB
string theory. These corrections to supergravity action come from
string theory tree-level scattering amplitude computations. $R^4$
term is the first higher order gravity correction terms in IIB
action. We may focus on this correction term. It corresponds to an
additional term in the action (\ref{iibaction})
\begin{eqnarray}
I &=& \frac{1}{16\pi G_{5}}\int d^{5}x \sqrt{-det\ g} \left[R-2
\Lambda+\gamma\mathcal{L}_{R^4}\right],
\end{eqnarray}
where $\gamma=\frac18 \zeta(3) (\alpha')^3$, and the correction
term in the Lagrangian density is of the form
\begin{eqnarray} \label{weyl}
\mathcal{L}_{R^4}&=&
C^{hmnk}C_{pmnq}{C_h}^{rsp}{C^q}_{rsk}+\frac12
C^{hkmn}C_{pqmn}{C_h}^{rsp}{C^q}_{rsk},
\end{eqnarray}
where $C_{pqmn}$ is the Weyl tensor.\\
First we calculate the entropy of Schwarzschild-AdS black hole
with $R^4$ correction. Using the metric (\ref{5dmetric}) one can
compute $\mathcal{L}_{R^4}$
\begin{equation}
\mathcal{L}_{R^4}=\frac{180r_0^{16}}{r^{16}l_{AdS}^8}.
\end{equation}
In this case, the modified metric reads\cite{gubser}
\begin{equation}
ds^2=-e^{2\lambda}dt^2+e^{2\nu}dr^2+\frac{r^2}{l_{AdS}^2}d\vec{x}^2,
\end{equation}
where
\begin{eqnarray}
e^{2\lambda}&=&\frac{r^2}{l_{AdS}^2}\left(1-\frac{r_0^4}{r^4}\right)\left[1-\frac{15\gamma
r_0^4}{l_{AdS}^6 r^{12}}(5r^8+5r^4r_0^4-3r_0^8)\right],\\
e^{2\nu}&=&\frac{r^2}{l_{AdS}^2}\left(1-\frac{r_0^4}{r^4}\right)^{-1}\left[1+\frac{15\gamma
r_0^4}{l_{AdS}^6 r^{12}}(5r^8+5r^4r_0^4-19r_0^8)\right].
\end{eqnarray}
Therefore the correction term to the entropy function reads
\begin{equation}
\mathbf{f}_{1BH}=-\frac{30\gamma V_3}{4\pi
G_5}\frac{r_0^{12}}{r^{9}l_{AdS}^{11}}\left(8+\frac{3r_0^4}{r^4}\right).
\end{equation}
The $R^4$ correction does not change the metric of the AdS
background\footnote{Generally speaking, to obtain the modified
metric, one should first variate the action and obtain the
equations of motion, which take the form
$$
R_{\mu\nu}-\frac12 g_{\mu\nu}R+\Lambda g_{\mu\nu}=\gamma
T_{\mu\nu}^{eff},
$$ where $T_{\mu\nu}^{eff}$ are constructed by Weyl tensor in present case. The
modified metric is then given by solving these equations of motion.
To the first order, we substitute the unperturbative metric into
R.H.S of the equations. For pure AdS and dS metric, it is not
difficult to check that all components of the Weyl tensor and hence
$T_{\mu\nu}^{eff}$ are vanishing. This implies the AdS metric is
unchanged when $R^4$ correction is included. This is also confirmed
in \cite{gubser}, where they calculate the black hole entropy using
Euclidean approach and claim that the action of AdS background $I_0$
is unchanged when $R^4$ is considered.
 }, so the correction of AdS entropy
function is vanishing, \textit{i.e.}, $\mathbf{f}_{1AdS}=0$. After
letting $R\rightarrow \infty$ we find the expression of the free
energy
\begin{eqnarray}
F&=&\int_{r_0}^{R}{\bf f}_{BH}-\left [\left(
    {g^{BH}_{tt} \over g^{AdS}_{tt}}
\right)^{1/2} \right]_{r=R}\int_{0}^{R}{\bf f}_{AdS},\\
&\simeq& -\frac{V_3 r_0^4}{16 \pi G_5
l_{AdS}^5}(1+\frac{75\gamma}{l_{AdS}^6}).
\end{eqnarray}
On the other hand, the temperature of the hole is
\begin{equation}
T=\frac{\kappa}{4\pi}=\left[\sqrt{g^{rr}}\frac{d}{dr}\sqrt{g_{tt}}\right]_{r=r_0}=\frac{r_0}{\pi
l_{AdS}^2}\left(1+\frac{15\gamma}{l_{AdS}^6}\right).
\end{equation}
Direct computation shows the corrected entropy of
Schwarzschild-AdS black hole with $R^4$ corrections is
\begin{eqnarray}
S&=&-\frac{\partial F}{\partial T}\\
&=& \frac{V_3 r_0^3}{4 G_5 l_{AdS}^3}
\left(1+\frac{60\gamma}{l_{AdS}^6} \right).
\end{eqnarray}
This is exactly the entropy of this case as shown in \cite{dutta}.

Again one can obtain the entropy of the Schwarzschild-dS black
hole using the same procedures above. Compared with the result for
the case without higher corrections, the entropy of the
Schwarzschild-dS black hole with $R^4$ corrections is different
from the entropy of the case for Schwarzschild-AdS black hole. In
present case, the entropy is
\begin{eqnarray}
S=\frac{V_3 r_0^3}{4 G_5 l_{dS}^3}
\left(1-\frac{60\gamma}{l_{dS}^6} \right).
\end{eqnarray}
\section{Conclusions}
In summary, we have discussed higher order corrections to black
hole entropy in dS and AdS spaces. We started from pure dS and AdS
spaces and we have found that even when the higher curvature
corrections are included, the corrected entropy for dS space can
still be calculated by using the equation $(TS)'=-\mathbf{f}$. The
 entropy function method and Wald's approach in calculating the entropies of dS
and AdS spaces agree with each other. As we have described, the
agreement between these two approaches can be understood, from
black hole thermodynamics \cite{gs}.

Although we do not know the general near horizon geometry of
non-extremal black holes in higher derivative gravity theory, our
results demonstrate that extending the entropy function formalism
to pure dS (AdS) space is safe. However, we do not find an
universal way to generalize the entropy function to more
complicated cases, such as non-extremal and rotating black holes.
For these cases one need first find that the symmetries of the
near horizon geometry for non-extremal and rotating black holes in
higher derivative gravity theory.

Note that the method used in this work is quite different from the
standard \textit{entropy function method } used in Ref.
\cite{sen,ag} in that the entropy is obtained without extremizing
the entropy function. Since there are no free parameters, such as
$v_i$ and $u_s$,  whose values should be determined by extremizing
the entropy function, the extremizing process is unnecessary. And
also, the near horizon limit is not done for (anti-) de Sitter
metric. The reason is because that for the  dS and  AdS metric, the
metric configuration is not so complicate as stringy black hole
metric.

\textbf{Acknowledgements}\\
The authors would like to thank G. W. Kang, S. P. Kim and M. I.
Park for their helpful comments at different stages of this work.
XHG is indebted to P. Zhang and S. Q. Wu for their hospitality
during his stay at SHAO and HNU. FWS would like to thank Y. Gong
and H. Yang for their hospitality during his stay at CQUPT and
UESTC.

\appendix


\section{de Sitter Entropy in $R^2$ gravity theories} \label{appendixA}

In this appendix we consider the entropy for pure de Sitter
spacetime with ``higher derivative'' gravity. The entropy of this
case has been shown to be given by (\ref{ds}) \be
TS=\int_{0}^{r_{H}}{\bf f}_{dS}dr,\ee where $r_H$ is the event
horizon of dS space. For $R^2$-gravity with cosmological constant,
the general action is given by
\begin{eqnarray}
S=\int d^{D}x \sqrt{- g}\left\{R-2\Lambda+a R^2+b
R_{\mu\nu}R^{\mu\nu}+cR_{\mu\nu\rho\sigma}R^{\mu\nu\rho\sigma}\right\}.
\end{eqnarray}
Variation over the metric $g_{\mu\nu}$ yields the equations of
motion as shown in \cite{cvetic}
\begin{eqnarray}
&&R^{\mu\nu}-{1 \over 2}g^{\mu\nu}\left(a R^2 + b R_{\mu\nu}
R^{\mu\nu} + c R_{\mu\nu\xi\sigma} R^{\mu\nu\xi\sigma} + R -
2\Lambda \right) \nonumber \\&& - a\left(-2RR^{\mu\nu} +
\nabla^\mu\nabla^\nu R +\nabla^\nu\nabla^\mu R
 - 2g^{\mu\nu} \nabla_\rho\nabla^\rho R \right) \nonumber\\
&& - b\left\{ {1 \over 2} \left(\nabla^\mu \nabla^\nu R +
\nabla^\nu \nabla^\mu R\right)
 - 2 R^{\mu\rho\nu\sigma} R_{\rho\sigma}
 - \Box R^{\mu\nu} - {1 \over 2} g^{\mu\nu}\Box R \right\}
 \nonumber\\
&&\nonumber - c\left(-2R^{\mu\rho\sigma\tau}R^\nu_{\
\rho\sigma\tau}
 - 4 \Box R^{\mu\nu} + \nabla^\mu \nabla^\nu R
+ \nabla^\nu \nabla^\mu R \right.\\&& \left. - 4
R^{\mu\rho\nu\sigma} R_{\rho\sigma} + 4 R^\mu_{\ \rho}
R^{\nu\rho}\right)=0.\label{eom}
\end{eqnarray}
We may expect that the modified metric of dS space takes the form
\begin{equation}\label{dsmetric}
ds^2=-\left[1-\frac{r^2}{l^2}(1+C)\right]dt^2+\left[1-\frac{r^2}{l^2}(1+C)\right]^{-1}dr^2+r^2
d\Omega_{D-2}^2,
\end{equation}
where $C$ is a constant which should be determined later. With
this metric ansatz, we obtain an equation of $C$ by substituting
the metric into Eqs. (\ref{eom})
\begin{eqnarray}
&&\nonumber[D(D-1)a+(D-1)b+2c](D-4)C^2+\{2[D(D-1)a+(D-1)b+2c](D-4)\\&&+(D-2)l^2\}C+[D(D-1)a+(D-1)b+2c](D-4)=0.
\end{eqnarray}
Straightforward computation gives
\begin{eqnarray}
\nonumber
C&=&-1-\frac{(D-2)l^2}{2[D(D-1)a+(D-1)b+2c](D-4)}\cdot\\&&\left\{1\mp
\sqrt{1+\frac{4[D(D-1)a+(D-1)b+2c](D-4)}{(D-2)l^2}}\right\}.
\end{eqnarray}
In particular, as we consider the Gauss-Bonnet combination,
\textit{i.e.}, $a=c,b=-4c$, we have
\begin{equation}
C=-1-\frac{l^2}{2(D-3)(D-4)c}\left\{1\mp
\sqrt{1+\frac{4(D-3)(D-4)c}{l^2}}\right\},
\end{equation}
which agrees with the result obtained in \cite{deser}.

Now it is safe to claim that \emph{dS space in $R^2$ gravity
theories has a modified metric in the form of (\ref{dsmetric})}.
In the following we will show that one can obtain the dS entropy
in any gravity theories with modified metric in the form of
(\ref{dsmetric}) without knowing the corrected metric. In order to
have a clear picture, we rewrite the metric (\ref{dsmetric}) as
\begin{equation}\label{dsmetric1}
ds^2=-\left[1-\frac{r^2}{l^2}(1+\gamma
\tilde{C})\right]dt^2+\left[1-\frac{r^2}{l^2}(1+\gamma
\tilde{C})\right]^{-1}dr^2+r^2 d\Omega_{D-2}^2,
\end{equation}
where $\gamma$ is the coupling strength, denoting $a$, $b$, $c$
for $R^2$ gravity theory, and $\tilde{C}$ is any constant. In any
higher derivative gravity theories, it is safe to expand $T$, $S$
and $\mathbf{f}$ in the following way
\begin{eqnarray}T=T_0+\gamma
T_1+\mathcal{O}(\gamma^2),\\
S=S_0+\gamma S_1+\mathcal{O}(\gamma^2),\\
{\bf f}={\bf f}_{0}+\gamma{\bf f}_{1}+\mathcal{O}(\gamma^2).
\end{eqnarray}
So from (\ref{ds}) we have
\begin{eqnarray}
T_0S_0+\gamma(T_0S_1+T_1S_0)=\int_0^{r_H^{(0)}}{\bf
f}_0dr+\int_{r_H^{(0)}}^{r_H^{(0)}+\gamma r_H^{(1)}}{\bf
f}_{0}dr+\gamma\int_0^{r_H^{(0)}}{\bf f}_1dr,
\end{eqnarray}
where $r_H^{(0)}$ corresponds to the horizon of dS space in the
case without the higher derivative, while $r_H^{(1)}$ is the first
order correction to the horizon when higher derivative is
considered. This implies the horizon for higher derivative gravity
is $r_H=r_H^{(0)}+\gamma r_H^{(1)}+\mathcal{O}(\gamma^2)$. In
particular, if the modified metric has the form as shown in
(\ref{dsmetric1}), we have $r_H^{(0)}=l$ and $r_H^{(1)}=-\frac{l
\tilde{C}}{2}$. Then $T_1$ can be evaluated as \be \label{T1}
T_1=\frac{r_H^{(1)}}{2\pi l^2}.\ee Note that $\mathbf{f}_1$
includes two parts: one comes from the Hilbert-Einstein action
(\textit{i.e.}, the Ricci scalar) evaluated by the modified metric
(\ref{dsmetric1}), here we denote it by $\mathbf{f}_1^{(1)}$; the
other corresponds to the $R^2$ gravity term calculated by
unperturbative metric, we use $\mathbf{f}_1^{(0)}$ to distinguish
it from the first one.

 For de Sitter spacetime, we have
$$S_0=\frac{l^{D-2}A_{D-2}}{4G_D},\ \ {\bf
f}_0=-\frac{(D-1)A_{D-2}r^{D-2}}{8\pi G_Dl^2}, \ \ {\bf
f}_1^{(1)}=-\frac{D(D-1)\tilde{C}A_{D-2}r^{D-2}}{16\pi G_D l^2}.$$
It is not difficult to show
\begin{eqnarray}
\gamma T_1S_0=\int_{r_H^{(0)}}^{r_H^{(0)}+\gamma r_H^{(1)}}{\bf
f}_{0}dr+\gamma\int_0^{r_H^{(0)}}{\bf f}_1^{(1)}dr.
\end{eqnarray}
Therefore Eq. (\ref{dsentropy}) shows that the entropy for dS
space in higher derivative gravity theory can be computed by
\begin{eqnarray}\label{dsentropy1}
S=S_0+\gamma S_1=\frac1{T_0}\left\{\int_0^{r_H^{(0)}}{\bf
f}_0dr+\gamma\int_0^{r_H^{(0)}}{\bf f}_1^{(0)}dr\right\}.
\end{eqnarray}

\bibliographystyle{plain}

\end{document}